\documentclass[12pt]{article}

\usepackage{url}
\usepackage{graphicx}
\usepackage{algorithm}
\usepackage{array}
\usepackage{amsmath}

\advance\topmargin by -0.95in
\begin{document}

\title{Optimizing Apportionment of Redundancies in Hierarchical RAID}
\author{Alexander Thomasian \\
Thomasian \& Associates     \\
Pleasantville, NY 10570 USA \\
alexthomasian@gmail.com}
\date{}
\maketitle

\begin{abstract}
Large disk arrays are organized into storage nodes - SNs or bricks 
ith their own cashed RAID controller for multiple disks. 
Erasure coding at SN level is attained via parity or Reed-Solomon codes.
Hierarchical RAID - HRAID provides an additional level of coding across SNs,
e.g., check strips P, Q at intra-SN level and R at the inter-SN level.
Failed disks and SNs are not replaced and rebuild is accomplished by restriping,
e.g., overwriting P and Q for disk failures and R for an SN failure.
For a given total redundancy level we use an approximate reliability analysis method
and Monte-Carlo simulation to explore the better apportionment of check blocks for intra- vs inter-SN redundancy.
Our study indicates that a higher MTTDL - Mean-Time-to-Data-Loss  
is attained by associating higher reliability at intra-SN level rather than inter-SN level,
which is contrary to that of an IBM study.
\end{abstract}


\section{Introduction to Hierarchical RAID - HRAID}\label{sec:intro20}

The five level classification of {\it Redundant Array of Independent Disks - RAID} 
introduced in \cite{PaGK88} was extended by RAID0 and RAID6 in \cite{Che+94}.
In this study we consider redundancy methods based on erasure coding,
rather than mirroring, which is inefficient in that it doubles disk space requirements.
RAID$(4+\ell), \ell \geq 1 >$ which is based on erasure coding utilizes 
the capacity equivalent of $\ell$ check disks to tolerate 
as many disk failures with {\it Maximum Distance Separable - MDS} codes \cite{MaSl77}.

Striping balances disk loads by partitioning data files into fixed size strips, 
which are placed round-robin across the disks.
Strips in a row constitute a stripe.
One strip per stripe is dedicated to parity in RAID5, two check strips in RAID6, 
Parity update loads are balanced by placing check strips in right to left diagonals \cite{Che+94}.

RAID performance is a primary issue in {\it OnLine Transaction Processing - OLTP},
which generates accesses to small randomly placed disk blocks.
A high positioning overhead (seek time plus latency) is incurred,
while the transfer time for small blocks is negligible.
Updating small blocks in RAID5 requires four disk accesses,
since given a new data block $d^{new}$,
we need to read the old data block $d^{old}$, unless it is cached,
read the corresponding parity block ($p^{old}$) to compute  
$d_{diff} = d_{old}  \oplus d_{new}$ and $p^{new} = p^{old} \oplus d^{diff}$/ 
Unless $d^{old}$ and $p_{old}$ are cached four disk accesses 
are required to update a single data block,
which is known as small write penalty \cite{Che+94}.
Given that the mean disk access for reads and writes is $\bar{x}_d$
and the fraction of read and (logical) write requests is $f_r$ and $f_w=1 - f_r$, 
then the average cost per disk access is $\bar{x}_{avg}= [f_r + 2 f_w (\ell+1)] x_d$.

{\it Hierarchical RAID - HRAID} was a proposal 
to apply the RAID paradigm at two levels \cite{Thom06b}.
IBM's Icecube is a similar proposal \cite{Wil+06}, 
which led to a prototype at a startup, but not a product. 
We consider an HRAID with $N$ {\it Storage Nodes - SNs} with $M$ disks per SN.
Each SN is a RAID0 or RAID$(4+\ell), 1 \leq \ell 3$ storage system with $\ell$ check strips per stripe.
Each SN is an $\ell$DFT $0 \leq \ell \leq 3$,
which can mask the failure of $\ell$ out of $M$ disks per SN.
HRAID$(k/\ell)$ extends the RAID paradigm to mask the failures $k$ out of $N$ SNs.
Up to $k$ SN failures can be tolerated by providing $0 \leq k \leq 3$ strips per stripe at each SN.
The average cost per disk access is then $x_{avg}= [f_r + 2 f_w (k+1)(\ell+1)] x_d$,
but there is the cost of transmitting $d^{diff}$.

\section{Intra- \& Internode Coding in HRAID}\label{sec:coding}

There are $N$ SNs and $M$ disks per SN.
Internode (resp. intranode) data protection is achieved via internode (resp. intranode) check strips.
Intranode check strips are computed over all strips at an SN, including internode check strips.
Internode check strips are computed over data strips at the same position at the remaining SNs,
but not intranode check strips.
In HRAID$k/\ell$ the coding across the $N$ SNs is $k${\it NFT},
while the coding at each SN with $M$ disks is $\ell${\it DFT}.


\begin{figure*}
\begin{tiny}
\begin{tabular}{|c|c|c|c||c|c|c|c||c|c|c|c||c|c|c|c|}\hline
\multicolumn{4}{|c||}{ Node 1} & \multicolumn{4}{c||}{ Node 2}
& \multicolumn{4}{c||}{ Node 3} & \multicolumn{4}{c|}{ Node 4} \\ \hline \hline
$~D_{1,1}^1$ & $~D_{1,2}^1$ & $~P_{1,3}^1$ & $~Q_{1,4}^1$ &
$~D_{1,1}^2$ & $~P_{1,2}^2$ & $~Q_{1,3}^2$ & $~D_{1,4}^2$ &
$~P_{1,1}^3$ & $~Q_{1,2}^3$ & $~D_{1,3}^3$ & $~D_{1,4}^3$ &
$~Q_{1,1}^4$ & $~D_{1,2}^4$ & $~D_{1,3}^4$ & $~P_{1,4}^4$ \\ \hline
$~D_{2,1}^1$ & $~P_{2,2}^1$ & $~Q_{2,3}^1$ & $~D_{2,4}^1$
& $~P_{2,1}^2$ & $~Q_{2,2}^2$ &$~D_{2,3}^2$ & $~D_{2,4}^2$
& $~Q_{2,1}^3$ & $~D_{2,2}^3$ & $~D_{2,3}^3$ & $~P_{2,4}^3$ &
$~D_{2,1}^4$ & $~D_{2,2}^4$ & $~P_{2,3}^4$ & $~Q_{2,4}^4$ \\ \hline
$~P_{3,1}^1$ & $~Q_{3,2}^1$ & $~D_{3,3}^1$ &  $~D_{3,4}^1$ &
$~Q_{3,1}^2$ & $~D_{3,2}^2$ & $~D_{3,3}^2$ & $~P_{3,4}^2$ &
$~D_{3,1}^3$ & $~D_{3,2}^3$ & $~P_{3,3}^3$ & $~Q_{3,4}^3$ &
$~D_{3,1}^4$ & $~P_{3,2}^4$ & $~Q_{3,3}^4$ & $~D_{3,4}^4$ \\ \hline
$~Q_{4,1}^1$ & $~D_{4,2}^1$ & $~D_{4,3}^1$ & $~P_{4,4}^1$ &
$~D_{4,1}^2$ & $~D_{4,2}^2$ & $~P_{4,3}^2$ & $~Q_{4,4}^2$ &
$~D_{4,1}^3$ & $~P_{4,2}^3$ & $~Q_{4,3}^3$ & $~D_{4,4}^3$ &
$~P_{4,1}^4$ & $~Q_{4,2}^4$ & $~D_{4,3}^4$ & $~D_{4,4}^4$ \\ \hline
\end{tabular}
\end{tiny}
\caption{\label{fig:hraid}A $4 \times 4$ HRAID1/1 with $N=4$ nodes and $M=4$ disks per node.
Only the first $M=4$ stripes are shown in the figure.}
\end{figure*}

The HRAID$k/\ell$ data layout with $N=M$ dedicates $k+\ell$ check strips per $M$ strips on an SN.
Check strips follow a left symmetric layouts as in RAID5 \cite{Che+94}.
Strips are shifted from SN to SN and on a per row basis at each SN.
The redundancy level is given as $(k+\ell)/M$,
While MDS - Maximum Distance Separable codes are used at both levels \cite{ThBl09},
the overall code is not MDS.
Given that HRAID$k/\ell$ tolerates $k$ SN failures and $\ell$ disk failures per SN
the maximum number of disk failures that can be tolerated is:
$$d_{max} = k \times M + (N-k)\times \ell = N (k+\ell) - k \ell,\mbox{ where }M=N$$.
Given that there are $N(k+\ell)$ check disks even in the best case fewer disk failures can be tolerated,
so that the overall code is not MDS.
It should be emphasized that MDS provides protection for all disk failure configurations,
while this is a best case scenario.

Inter-SN check blocks allow the recovery of missing blocks if intra-SN recovery fails.
This results in a significant improvement in {\it Mean Time To Data Loss - MTTDL} 
with respect to HRAID$0/\ell$ with no inter-SN check blocks.

When a disk fails performance improvements is attained via restriping,
i.e., overwriting check strips with data strips or the data on a whole SN. 
Starting with RAID7 with check blocks P, Q, R restriping 
which overwrites the blocks in the order R, Q, P results in         
$\mbox{RAID7} \rightarrow \mbox{RAID6} \rightarrow \mbox{RAID5} \rightarrow \mbox{RAID0}$       \newline
An SN failure occurs due to its controller failures or 
when the number of its failed disks at an SN exceeds $\ell$.

\section{Shortcut Reliability Analysis of HRAID}\label{sec:SRA2}

Let $r = 1 - \epsilon$ denote the reliability of each disk, where $\epsilon \ll 1$ is the disk unreliability.
Let $R_{\ell}$ denote the reliability of RAID$(4+\ell)$ with $R_0$ for RAID0,
e.g., in the case of RAID5 $\ell=1$ which can tolerate one disk failure \cite{Thom06a}:
\vspace{-2mm}
\[
R_1 = r^M + N (1-r) r^{M - 1}  = (1-\epsilon)^M + M \epsilon (1-\epsilon)^{M-1}
\approx 1- {M \choose 2} \epsilon^2  + 2 {M \choose 3} \epsilon^3 -\ldots
\]

It is shown in \cite{Thom06a} that the smallest power $n$ in $\epsilon^n$ of the polynomial
determines the minimum number of disk failures leading to data loss, which is two in this case.
In other words RAID5 can tolerate a single disk failure.
The approximate reliability equation for RAID$(\ell+4), \ell \geq 1$ obtained by induction is:
\vspace{-2mm}
$$
R_{\ell} \approx 1 - 
\binom{N}{\ell+1} \epsilon^{\ell+1}  + (\ell+1) \binom{N}{\ell+2} \epsilon^{\ell+2} - \ldots
$$
Note that the first term indicating the probability of data loss
due to $\ell+1$ disk failures is subtracted from one.

Given that $R_{k/\ell}(N,M)$ denote the reliability of HRAID$k/\ell$
with $N$ nodes and $M$ disks as affected by disk failures only,
the reliability of HRAID$1/0 (N,M)$ and HRAID$0/1 (N,M)$ can be expressed
by first noting that $R_1 = r^M + M(1-r)r^{M-1}$ and $R_0=r^M$:
\vspace{-2mm}
\begin{eqnarray*}
R_{0/1} (N,M) = (R_1)^N \approx 1- \frac{NM(M-1)}{2}\epsilon^2.
\end{eqnarray*}
\vspace{-2mm}
\begin{eqnarray*}
R_{1/0} (N,M) = (R_0)^N + N(1-R_0)(R_0)^{N-1} \approx 1 - \frac{N(N-1)M^2}{2} \epsilon^2.
\end{eqnarray*}
It follows that $d_{min}= (k+1)(\ell+1)=2$ disk failures may lead to data loss in both cases, 
but $R_{0/1} > R_{1/0}$.

It follows from the expressions for approximate reliability
for HRAID$1/2$ and HRAID$2/1$ that both can fail with $(k+1)(\ell+1)=6$ disk failures
\[
R_{1/2} (N,M) = R_2^N  + N (1 - R_2 )R_2^{N-1}
\approx 1- \frac{N(N-1)M^2 (M-1)^2 (M-2)^2}{72} \epsilon^6 + \ldots
\]
\[
R_{2/1} (N,M) = R_1^N + N(1-R_1)R_1^{N-1} + {N \choose 2} (1-R_1)^2 R_1^{N-2}
\approx 1 - \frac{N(N-1)(N-2)M^3 (M-1)^3}{24} \epsilon^6 + \ldots
\]
and that HRAID$1/2$ is more reliable than HRAID$2/1$,
since $R_{1/2} (N,M) > R_{2/1} (N,M)$ implies $N > 2+ (M-2)^2 / (3M(M-1))$,
which is true for reasonable values of $N$.

To determine the probability that HRAID1/2 encounters data loss with the sixth disk failure,
consider the configuration with one failed node due to three failed disks and another node with two failed disks,
The probability of a disk failure at this node is $p_{1/2} = (M-2)/D_S$ with $D_S = (N-2)M + M-2$.
In the case of HRAID2/1 array consider two nodes with two failed disks each,
so that their data can only be reconstructed via the internode check code.
Data loss occurs when a third node, which already has a failed disk, encounters a second disk failure.
The probability of this event is $p_{2/1} = (M-1)/D_S$.
The inequality $p_{1,2} < p_{2,1}$ leads to $ N+M > 4$, which is always true.

Monte Carlo simulation was used in \cite{ThTH12} to determine the MTTDL.
Based on the data provided in \cite{Gibs92} the time to disk failure 
is assumed to be exponentially distributed \cite{Triv01}
which can be specified by a single parameter: {\it Mean Time To Failure - MTTF}
million hours or equivalently failure rate $\delta=10^{-6}$ 
It follows that the total failure rate is 
the sum of the failure rates of surviving components .
A more detailed discussion of reliability analysis appears in \cite{Thom21}.
The assumption that controllers do not fail (failure rate $\gamma=0$)
leads to Table 4 in \cite{ThTH12}. 

\begin{table}[h]
\begin{center}
\caption{HRAID$K/\ell$ MTTDL in thousands of hours for $N=M=12$ and disk MTTF=$10^6$ hours.} 
\begin{tabular}{|c|c|c|c|c|} \hline
               &$k=0$  &$k=1$ &$k=2$ &$k=3$     \\ \hline
$\ell=0$       &6.9    &14.6  &23    &32        \\
$\ell=1$       &36.9   &58.9  &78.4  &97.7     \\
$\ell=2$       &118.9  &118.8 &148.7 &176.8     \\
$\ell=3$       &139.6  &191.5  &231.8 &268.1    \\
\hline
\end{tabular}
\end{center}
\end{table}

It is easy to see that higher redundancy at lower level is preferable,
but this conclusion is true to a lesser degree 
for controller failure rates equalling or exceeding those of disks $\gamma=1,2,3 \times 10^{-6}$.
For the specific target chosen for Icecube in \cite{RaHG11}
HRAID2/1 or HRAID3/0 meet the reliability  requirement,
while our analysis contradicts this conclusion.
The hierarchical reliability analysis method developed in \cite{RaHG11} 
is approximate and require validation, via simulation similar to the one used in \cite{ThTH12}.  

\subsection*{Acknowledgments}

In carrying out the simulation at {\it Shenzhen Institutes of Advanced Technology - SIAT}
I was assisted by Ms Yujie Tang (now at Algoma Univ.). 
Mr Yang Hu (now at Tencent) corrected a bug in the simulation.


\begin{thebibliography}{10}

\bibitem{Che+94}
P. M. Chen, E. K. Lee, G. A. Gibson, R. H. Katz, D. A. Patterson:
RAID: High-performance, reliable secondary storage. 
{\it ACM Comput. Surv. 26}(2): 145-185 (1994).

\bibitem{Gibs92}
G. A. Gibson:
Redundant Disk Arrays: Reliable, Parallel Secondary Storage.
MIT Press,  1992

\bibitem{MaSl77}
F. J. MacWilliams, N.J. Sloane.
The Theory of Error-Correcting Codes.
North-Holland 1977.

\bibitem{PaGK88}
D. A. Patterson, G. A. Gibson, R. H. Katz:
A case for redundant arrays of inexpensive disks (RAID). 
{\it Proc. ACM SIGMOD Conf.} 1988: 109-116

\bibitem{RaHG11}
K. K. Rao, J. L. Hafner, R. A. Golding:
Reliability for Networked Storage Nodes. 
IEEE Trans. Dependable Secure Computing 8(3): 404-418 (2011)

\bibitem{Thom06a}
A. Thomasian:
Shortcut method for reliability comparisons in RAID. 
{\it J. Systems Software 79}(11): 1599-1605 (2006).

\bibitem{Thom06b}
A. Thomasian:
Multi-level RAID for very large disk arrays. 
{\it ACM SIGMETRICS Perform. Evaluation Rev. 33}(4): 17-22 (2006).

\bibitem{ThBl09}
A. Thomasian, M. Blaum:
Higher reliability redundant disk arrays: Organization, operation, and coding. 
{\it ACM Trans. Storage 5(3)}: 7:1-7:59 (2009)

\bibitem{ThTH12}
A. Thomasian, Y. Tang, Y. Hu:
Hierarchical RAID: Design, performance, reliability, and recovery. 
{\it J. Parallel Distributed Computing 72}(12): 1753-1769 (2012).

\bibitem{Thom21}
A. Thomasian:
Storage Systems: Organization, Performance, Coding, Reliability, and Their Data Processing, 
Morgan Kaufmann / Elsevier 2021,

\bibitem{Triv01}
K. S. Trivedi.
Probability and Statistics with Reliability, Queueing, and Computer Science Applications, 2nd Ed.
Wiley 2001.

\bibitem{Wil+06}
Winfried W. Wilcke et al.
IBM Intelligent Bricks project - Petabytes and beyond. 
IBM J. Research \& Development 50(2-3): 181-198 (2006)

\end{thebibliography}
\end{document}